\documentstyle[12pt]{article}
\normalsize

\def\Hat#1{\rlap{\kern.10em$\widehat{\phantom G}$}#1}
\def\HAt#1{\rlap{\kern.05em$\widehat{\phantom G}$}#1}

\def\cap#1{\rlap{\kern.1em$\widehat{\phantom{G\vrule height.8em}}$}#1{}}
\def\Cap#1{\rlap{\kern.05em$\widehat{\phantom{G\vrule height.8em}}$}#1{}}

\newcounter{sxn}

\newcounter{axn}

\def\br{\begin{thebibliography}{abc}}
\def\er{\end{thebibliography}}

\date{}

\begin{document}
\bibliographystyle{unsrt}
\footskip 1.0cm
\thispagestyle{empty}
\setcounter{page}{0}
\begin{flushright}
SU-4240-569\\
February, 1994 \\
\end{flushright}
\vspace{10mm}

\centerline {\LARGE Fractional Quantum Hall Effect}
\vspace{5mm}
\centerline {\LARGE From Anomalies In WZNW Model}
\vspace*{15mm}
\centerline {\large  L. Chandar}

\vspace*{5mm}
\centerline {\it Department of Physics, Syracuse University,}
\centerline {\it Syracuse, NY 13244-1130, U.S.A.}
\vspace*{15mm}
%\baselinestretch{2.0}
\normalsize
\centerline {\bf Abstract}
\vspace*{5mm}

An approach to understand
Fractional Quantum Hall Effect (FQHE) using anomalies is studied in this paper.
More specifically,
this is done by looking at the anomaly in the current conservation equation of
a WZNW theory describing fields living at the edge of the two dimensional
Hall sample.  This WZNW theory itself comes from the non-Abelian bosonisation
of fermions living at the edge.  It is shown that this model can describe both
integer and fractional quantization of conductivities in a unified manner.

\newpage

\baselineskip=24pt
\setcounter{page}{1}
\newcommand{\be}{\begin{equation}}
\newcommand{\ee}{\end{equation}}

It has been known for a long time
that the quantum Hall liquid is an incompressible
fluid.  Since incompressible fluids are in general characterised by degrees of
freedom at the boundary \cite{1,3,17,38}, so it is for the quantum Hall system
too.  In fact,
Halperin was the first to point this fact out \cite{2} in the context of
integral quantum Hall effect (IQHE) when he showed that the quantization of
conductance can be
directly obtained by studying the behaviour of the edge currents.  Edge states
have subsequently played a significant role in attempts to understand different
aspects of both IQHE and FQHE (See \cite{1,3,17} and references therein for
more details).

There have been two kinds of approaches to understand these edge states.  One
of them has used Chern-Simons type effective actions  in
the 2+1
dimensional Hall system, where it is
known that all observables reside only at the edge.  The other approach uses
a 1+1 dimensional field theory which describes directly these
edge observables.  That these two approaches are equivalent has been shown in
references \cite{18,7} (see also \cite{37} and references therein).

The advantage of using the second approach is that  the standard
bosonisation techniques apply in 1+1 dimensions and so we can freely go
from a fermionic description to a bosonic one.

For example, a free fermionic theory with one fermion is equivalent to a free
bosonic theory also with one boson.  If the fermion were chiral, so would be
the
corresponding boson.  If, on the other hand, the free fermionic theory had $N$
fermions, then there is a $U(N)$ (or $O(2N)$) symmetry leaving the original
action invariant. Here, the corresponding bosonised theory should also have
this symmetry.  As is well known, this argument leads to the standard
non-Abelian bosonised
action that Witten first wrote down \cite{8}.

The physical motivation for considering the case with many fermions rather than
just one is the following.  In a model proposed by Jain
\cite{9}, an individual electron is assumed to be composed of $N$ ``partons''.
Such a model (with some additional assumptions), interestingly enough, is able
to explain a number of features observed experimentally in FQHE, including the
actual fractions observed and their relative stability as compared to other
observed fractions.  In such a case, the associated edge fermions should also
be $N$ in number.  Another reason could be that the edge fermions of each
Landau level appear independently \cite{10} so that in a situation with $N$
Landau levels being filled, one has to consider also $N$ edge fermions.
Yet other physical reasons for considering a theory with $N$ fermions also
exist.  For example, we may have the requirement of a theory with a $U(N)$
symmetry because of the presence of such a symmetry mixing the particles and
quasiparticles of different levels in the hierarchical approach \cite{21,22}
used to understand the fractions observed in FQHE \cite{23}.

The non-Abelian bosonised action which describes the bosonised edge
fermions using group valued bosonic fields $g(x)$ is
\be
S = -\frac{n}{8\pi}\int _{\Sigma} Tr(g^{-1}\partial _{\mu}g)^{2} +
\frac{n}{12\pi}\int _{\Sigma ^{0}} Tr(g^{-1}dg)^{3}   \label{1}
\ee
where $Tr$ refers to the Trace taken in the fundamental representation,
$\Sigma$ is the space-time manifold (1+1 dimensional), while $\Sigma ^{0}$ is
such that $\partial \Sigma ^{0} =\Sigma $.

In the above, $n$ has to be an integer in order that the action gives rise to a
sensible quantum theory.  According to Witten, the above theory describes the
bosonised version of the free fermionic theory
with $N$ free fermions \cite{8} when $n$ is equal to 1 and only for this value
of $n$.  However, Witten also showed by a one-loop
calculation as well as by showing the equivalence with an exactly solvable
conformally invariant theory that the $\beta$ function for the above theory
vanishes identically for all integral $n$.  In fact, the following arbitrary
non-linear model with the WZNW term,
\be
S = -\frac{1}{2\lambda ^{2}}\int _{\Sigma}Tr(g^{-1}\partial _{\mu}g)^{2} +
\frac{n}{12\pi}\int _{\Sigma ^{0}}Tr (g^{-1}dg)^{3} \label{1.5}
\ee
flows into one with $\lambda ^{2} = \frac{4\pi}{n}$ for long distances.  This
means that for a class of interacting fermionic theories, the corresponding
bosonised theories will, to a good approximation, show features identical to
that of a theory described by (\ref{1}) for some integer $n$.  Moreover, this
approximation becomes
increasingly better at longer scales of lengths ({\it i.\@e.}\ in the limit of
large size of the Hall sample ).

The action (\ref{1.5}) when gauged with a $U(1)$ charge should correspond
therefore
to a situation where the fermions are coupled to electromagnetic field.
In addition edge states for large samples should be associated with $\lambda
^{2} =\frac{4\pi}{n}$, which is an infrared fixed point.

Before doing this gauging, let us first look at the Abelian free boson which is
equivalent to a single free fermion \cite{11}.  This analysis will also help in
knowing the essential differences that arise because of the non-Abelian nature
of the action (\ref{1}).  The action for this theory
is
\be
S = \frac{1}{2}\int \partial _{\mu}\phi \partial ^{\mu}\phi \label{2}
\ee
This action when gauged with a $U(1)$ charge is (where the gauging
is done in accordance with the transformation law
$e^{i\phi} \rightarrow e^{i(\phi -e\gamma )}$)
\be
S[\phi , A] = \frac{1}{2}\int (\partial _{\mu}\phi + eA_{\mu})^{2} -
\frac{1}{4k^{2}}\int F_{\mu\nu}F^{\mu\nu} \label{3}
\ee
As it stands the above theory is gauge invariant and so has no
anomalies.  But from the following argument \cite{1} (which is essentially a
particular case of the Callan-Harvey effect \cite{12}), we know that a
situation with Hall
conductivity corresponds to a situation with the electromagnetic current in
1+1 dimensions satisfying an anomalous divergence equation (see also
\cite{33}).

Consider a disc of radius $R$ and an electric field $E$ tangential to the
boundary everywhere (this can be arranged by having a constantly increasing
magnetic flux within the disc).
Then, if $\sigma _{H}$ denotes the Hall conductivity and $j_{r}$ the radial
current,
\begin{eqnarray}
j_{r} & = & \sigma _{H}E \label{4}  \\
\Rightarrow j_{r}(2\pi R) & = & \sigma _{H}E(2\pi R) \label{5}
\end{eqnarray}
implying that the rate of increase of charge at the boundary of the disc is
\be
\dot{Q} = 2\pi R\sigma _{H}E.\label{5.5}
\ee
But for the 1+1 dimensional current at the boundary of the disc,
\be
\partial _{\mu} j^{\mu} = \partial _{0} j^{0} - \partial _{1} j^{1} \label{6}
\ee
so that if the theory is anomalous with the usual Schwinger anomaly, then
\be
\partial _{0} j^{0} - \partial _{1} j^{1} = \alpha E\label{7}
\ee
where $\alpha$ is the anomaly coefficient and $E$ is the same as $F_{01}$.
{}From (\ref{6}) and (\ref{7}) it follows that
\begin{eqnarray}
\int _{\partial D} (\partial _{0} j^{0} - \partial _{1} j^{1}) & = & \alpha E
(2\pi R) \label{8}\\
\mbox{or \ \ }\int _{\partial D} \partial _{0} j^{0} & = & 2\pi R \alpha E
\label{9}
\end{eqnarray}
implying therefore that the rate of increase of charge at the boundary of the
disc is
\be
\dot{Q} = 2\pi R\alpha E. \label{10}
\ee
{}From (\ref{5.5}) and (\ref{10}), it follows that $\alpha = \sigma _{H}$ and
so a nonzero $\sigma _{H}$ requires an anomalous theory.

Thus the exact gauge invariance of the Abelian action (\ref{2}) is not
satisfactory since we
have to add an anomaly term to it by hand in order to approximate the physical
situation.

Without further physical input, there is no way to fix the strength of the
anomaly coefficient in the bosonic theory.  The crucial ingredient which fixes
this coefficient and also generalises to
the
non-Abelian case is the physical requirement of chirality which we have not yet
imposed on the fields and currents.  Such a requirement is physically needed
because a Hall
sample is placed in a large magnetic field perpendicular to the sample and so
the fields and currents excited are necessarily chiral.  As we will see below
this chirality constraint requires that there is an anomaly. In
addition it also
fixes the anomaly coefficient.  This is satisfying because it means that the
Hall conductivity is determined uniquely by chirality for a given theory.

With this chirality constraint,  the equations of motion of above theory
\begin{eqnarray}
\frac{1}{k^{2}}\partial _{\mu}F^{\mu\nu} +e(\partial ^{\nu}\phi +
eA^{\nu}) & = & 0 \nonumber \\
\partial _{\mu}(\partial ^{\mu}\phi + eA^{\mu}) & = & 0 \label{11}
\end{eqnarray}
have to be supplemented by (say) the equation
\be
D_{+}\phi \equiv D_{0}\phi + D_{1}\phi = 0 \label{12}
\ee
which is the condition that a left-moving $\phi$ has to satisfy.
But then
\begin{eqnarray}
0 =\partial _{-}D_{+}\phi & = & (\partial _{0} -\partial _{1})(D_{0}\phi
+D_{1}\phi ) \nonumber \\
& = & \partial _{\mu}D^{\mu}\phi + eE \nonumber \\
\mbox{or \ } \partial _{\mu}D^{\mu}\phi + eE & = &  0. \label{13}
\end{eqnarray}
Thus we see that the constraint (\ref{12}) leads to an equation inconsistent
with (\ref{11}).  So the action (\ref{3}) has to be modified to
\be
S' = \frac{1}{2}\int (\partial _{\mu}\phi + eA_{\mu})^{2} -
\frac{1}{4k^{2}}\int F_{\mu\nu}F^{\mu\nu} + e\int \phi \epsilon
_{\mu\nu}\partial ^{\mu}A^{\nu} \label{13.5}
\ee
in order that its equations of motion are compatible with imposition of
condition (\ref{12}).  The current $j_{\mu}$ for this action defined as
$\frac{\delta S_{matter}}{\delta A^{\mu}}$ is
\be
j_{\mu} = e[D_{\mu}\phi + \epsilon _{\mu\nu}\partial ^{\nu}\phi ] \label{14}
\ee
However this current is neither gauge invariant nor does it satisfy the
chirality condition $j_{+}\equiv j_{0} +j_{1}=0$.  Both these drawbacks are
resolved\footnote{I am thankful to A.P.Balachandran and B.Sathiapalan for
helping me resolve this issue.} by considering the current ${\cal J}_{\mu}$
got by ``covariantising'' (\ref{14}):
\be
{\cal J}_{\mu} = e[D_{\mu}\phi + \epsilon _{\mu\nu}D^{\nu}\phi ]. \label{14.5}
\ee
This current is gauge invariant by construction and it can be checked that
${\cal J}_{+}$ for this current is identically zero.  Thus this is the physical
current which describes edge excitations for the Hall system.  In a subsequent
paper \cite{13}, we will also justify this modification of the current by
considering the
contribution to the total current at the boundary coming from the theory in the
interior of the disc.

Using the equation of motion from the action (\ref{13.5}), we get
\be
\partial _{\mu}{\cal J}^{\mu} = -2e^{2}E. \label{14.6}
\ee
So the physical
picture we have is that if we require chirality, then anomaly, including the
coefficient of anomaly (and therefore the
Hall conductivity) is forced  on us (cf. \cite{13} for further results using
this approach).

If, on the other hand, we require this anomaly even without having to impose
chirality, clearly there is no natural way to obtain it in the Abelian theory.
This problem is circumvented by looking at the non-Abelian action (\ref{1}),
where, as is well known \cite{14,15} gauging by an arbitrary subgroup in
general
leads to anomalies.    In particular, we can then
suppress the modes of one chirality at the end if we only want to
work with chiral objects (as in a real Hall system).  In such a case, we
should then be careful whether
the anomaly coefficient as obtained by naive gauging matches the anomaly
coefficient as required by imposing chirality.  We will see below that these
indeed coincide under certain conditions.

Thus, if we gauge  (\ref{1}) with a chiral charge $Q$ which acts on
the group element $g$ by multiplication from the left as follows:
\be
g \rightarrow e^{ie\epsilon Q} g,\label{14.9}
\ee
the gauged action will be
\be
S[g, A] = -\frac{n}{8\pi}Tr\int (g^{-1}D_{\mu}g)^{2} +\frac{n}{12\pi}Tr\int
(g^{-1}dg)^{3}  -\frac{1}{4k^{2}}\int F_{\mu\nu}F^{\mu\nu} +
\frac{ien}{4\pi}Tr\int (AQ,dgg^{-1})\label{15}
\ee
where $D_{\mu} = \partial _{\mu} -ieA_{\mu}Q$  and $F_{\mu\nu}
=\partial _{\mu}A_{\nu} -\partial _{\nu}A_{\mu}$.

The above action of course is not completely gauge invariant (neither can any
further terms be added to cancel the terms that change under a gauge
transformation) and so the equations of motion of this action will have anomaly
terms:\footnote{In the following, we use the metric ($+,-$) and use the
convention that $\epsilon _{01}=-\epsilon ^{01}=1$.}
\begin{eqnarray}
\frac{1}{k^{2}}\partial _{\nu}F^{\nu\mu} & = & -j^{\mu} =
-\frac{ien}{4\pi}Tr[Q(D^{\mu}gg^{-1} - \epsilon ^{\mu\nu}\partial
_{\nu}gg^{-1})] \nonumber \\
\partial _{\mu}j^{\mu} & = & \frac{ne^{2}}{4\pi}ETr(Q^{2})\label{16}
\end{eqnarray}
For posterity, we will also note here the remaining equations that follow from
the above action (a particular case of which gives the last of equations
(\ref{16})):
\begin{eqnarray}
\partial _{\mu}(D^{\mu}gg^{-1} -\epsilon ^{\mu\nu}\partial _{\nu}gg^{-1})&=&
ie\epsilon ^{\mu\nu}(\partial _{\mu}A_{\nu}Q+A_{\mu}[\partial
_{\nu}gg^{-1},Q])\nonumber\\
\mbox{(or)\ \ } \partial _{\mu}(g^{-1}D^{\mu}g+\epsilon ^{\mu\nu}g^{-1}\partial
_{\nu}g)&=& ie\epsilon ^{\mu\nu}g^{-1}(\partial _{\mu}A_{\nu}Q+A_{\mu}[\partial
_{\nu}gg^{-1},Q])g \label{16.1}
\end{eqnarray}

As in the Abelian case, here too we see that $j^{\mu}$ (defined using
$\frac{\delta S_{matter}}{\delta A_{\mu}}$) is not gauge invariant.  We thus
``covariantise'' $j^{\mu}$ to get the physical gauge invariant current
\be
{\cal J}^{\mu} = \frac{ien}{4\pi}Tr[Q(D^{\mu}gg^{-1} - \epsilon
^{\mu\nu}D_{\nu}gg^{-1})]. \label{16.5}
\ee
Using (\ref{16}) we see that
\be
\partial _{\mu} {\cal J}^{\mu} = \frac{ne^{2}}{2\pi}ETr(Q^{2}).\label{16.8}
\ee

It should be emphasized here that the redefinition (\ref{16.5}) of the physical
current is crucial in giving the correct Hall factor of $\frac{e^{2}}{2\pi}$
in equation (\ref{16.8}) above.  With the current $j^{\mu}$, the anomaly would
have been wrong by a factor of 1/2 \cite{13}.

One observation here is that the simplest choice for the charge $Q$ is the one
which is
proportional to the identity (assuming that the $N$ fermions are identical so
that
each of them has the same charge) so that the gauge transformation law
(\ref{14.9}) is same
as multiplication by a phase.  But in this case, the action (\ref{15}) is gauge
invariant without the last term.  For any other $Q$, since the WZNW
term is not by itself gauge invariant, this last term is required to cancel
the change in the WZNW term.  Thus, if we interpret the physical $U(1)$ charge
$Q$ (which lies in the center of $U(N)$) as a limit of a sequence of charges
such
that only the limit lies in the center of $U(N)$,  then
the last term of (\ref{15}) (which is also the term which contributes to the
anomaly in the current conservation equation) is automatically required.
Slightly different arguments (by increasing the group to $U(N+1)$ and calling
the charge as that diagonal generator which is 1 on the first $N$ diagonal
entries and $-N$ at the last diagonal entry, so that when restricted to the
$U(N)$ subgroup, it is just like a multiplicative phase) justifying the
inclusion of such a  term in a
similar model (the Skyrme model describing baryons; see, for example \cite{16})
have also appeared in
the literature.   Even though these arguments  appear
quite contrived, there is the physical requirement of chirality that
we have
not yet imposed.  As we will show later, this physical requirement gives us the
same anomaly as the one obtained by the above arguments.

{}From equation (\ref{16.8}), we read off
\be
\sigma _{H} = \frac{ne^{2}}{2\pi}Tr(Q^{2})\label{17}
\ee
(By choosing the gauge transformation law to multiply from the right rather
than from the
left as in (\ref{14.9}), the opposite sign is obtained.)
It follows that for arbitrary integer $n$, each of which is a theory with
vanishing
$\beta$-function and therefore  a physical limiting theory for a class of
interacting fermionic theories in the limit of large radius, the $\sigma _{H}$
is always an integer multiple of the same basic conductivity.  Thus Integral
Quantum Hall Effect (IQHE) follows as a direct consequence of
starting with fermionic theories with different interactions so that they flow
into different infrared limits.  This result is quite interesting because of
its generality.  It says that the quantization of Hall conductivity does not
require any specific assumptions regarding the nature of interactions between
the electrons in the sample since the long wavelength limit of a class of
theories goes to one with a quantized Hall conductivity.

The case of greater interest for us is when we use the analog of Jain's model.
At the outset,
it may seem that formula (\ref{17}) only gives integer quantization and may
seem paradoxical because Jain's model was necessary precisely to explain
the fractional conductivity.  But this is not so, because of the free parameter
$N$ which is the number of partons which composed a given electron and which
enters in $Tr(Q^{2})$ in the formula (\ref{17}).
One of the obvious conditions to impose on $N$ is that it be odd in
order that the composite of these $N$ partons (each of which is assumed to be a
fermion) is also a fermion. One  other condition that needs to be imposed is
that
the total charge of the composite, which is the sum of the diagonal entries in
the fundamental representation of the charge matrix, be equal to  the charge of
the electron.  We may also require that in a direct product representation
obtained from $N$ fundamental representations the eigenvalues of the charge
matrix are all integer multiples of $e$, the electronic charge.

Since the electron itself has a given charge
$e$, the simplest choice for the charge matrix $Q$ is
$1\!\!\! \mbox{l}/N$ so that the charge of each parton is $1/N$ times $e$.
Thus
\be
\sigma _{H} = \frac{ne^{2}}{2\pi N} \label{18}
\ee
and therefore the allowed values of the Hall conductivity (\ref{18}) are, apart
from a constant factor,
fractions with odd denominators.  Clearly we could have obtained other sets of
values for $\sigma _{H}$ if the charge $Q$ is chosen  not to lie in the center
of $U(N)$ so that $Tr(Q^{2})$ is accordingly different.  Of course, the allowed
$Q$ are constrained by the requirements that the sum of the diagonal entries
is equal to $e$ and that in a higher dimensional representation obtained by
taking the direct product $N$ times of the fundamental representation, an
eigenvalue of the charge matrix is an integral multiple of $e$.

To arrive at the chiral theory (and for the true justification of the anomaly
term in (\ref{15}) for a central $U(1)$ charge),  we need to  impose on
the states the chirality constraint
\be
\frac{ien}{4\pi}Tr[Qg^{-1}D_{-}g]= 0 \label{19}
\ee
Of course, as before, we have to check the consistency of the above constraint
with the equations of motion (\ref{16.1}).  It turns out that in fact this
consistency is satisfied because of the extra anomaly piece
in the conservation equation, exactly as it happened in the Abelian case:
\begin{eqnarray}
0 & = &  \partial _{+} (\frac{ien}{4\pi})Tr[Qg^{-1}D_{-}g]\nonumber \\
\Rightarrow 0 & = & \partial _{+} (\frac{ien}{4\pi})Tr[Qg^{-1}\partial_{-}g]
+\frac{e^{2}n}{4\pi}\partial _{+}Tr[Qg^{-1}A_{-}Qg]\nonumber \\
\Rightarrow 0 & = & \frac{ien}{4\pi}Tr[Q(\partial
_{\mu}(g^{-1}D^{\mu}g+\epsilon ^{\mu\nu}g^{-1}\partial _{\nu}g)+ie\partial
_{\mu}(g^{-1}A^{\mu}Qg))]+\nonumber\\
&&\frac{e^{2}n}{4\pi}\partial
_{\mu}Tr[Qg^{-1}(A^{\mu}+\epsilon ^{\mu\nu}A_{\nu})Qg]\nonumber \\
\mbox{ (on using (\ref{16.1})) } \nonumber \\
\Rightarrow 0 & = & 0, \mbox{{\it i.\@e.} no inconsistency } \label{19.5}
\end{eqnarray}
Furthermore, the physical gauge invariant current ${\cal J}^{\mu}$ defined in
(\ref{16.5}) satisfies ${\cal J}_{-}\equiv {\cal J}_{0}-{\cal J}_{1} =0$ as an
identity.

As mentioned even earlier, it is interesting to note that the
coefficient of anomaly as obtained by these two seemingly different approaches
(one
enforcing chirality, the other by gauging a WZNW term)
coincide!  It is also worth noting here that if instead of the
chirality condition (\ref{19}), we imposed the condition
\be
\frac{ien}{4\pi}Tr[QD_{+}gg^{-1}]=0, \label{19.6}
\ee
we would not be able to satisfy the above consistency
with the equations of motion.  Thus on using (\ref{16}), we have that
\begin{eqnarray}
0 & = & \partial _{-} (\frac{ien}{4\pi})Tr[Q D_{+}gg^{-1}]
\nonumber\\
\Rightarrow 0 & = & \frac{ien}{4\pi}Tr[Q(\partial
_{\mu}(D^{\mu}gg^{-1}-\epsilon ^{\mu\nu}\partial _{\nu}gg^{-1})+ieQ\partial
_{\mu}A^{\mu})]+\frac{e^{2}n}{4\pi}Tr(Q^{2})\partial _{\mu}(A^{\mu}-\epsilon
^{\mu\nu}A_{\nu})\nonumber \\
\Rightarrow 0 & = & 0 +\frac{e^{2}n}{2\pi}ETr(Q^{2}) \label{19.7}
\end{eqnarray}
which is an inconsistency.

This asymmetry between the two
chiralities is due to the fact that the gauging
of (\ref{1}) was done by
multiplication from the left:
\begin{flushright}
$g \rightarrow e^{ie\epsilon Q} g$ \hspace{3.5in} (\ref{14.9})
\end{flushright}
If, instead, the gauging had been done by multiplication from the right, it is
the constraint (\ref{19.6}) which would have been consistent with the equations
of motion, whereas the constraint (\ref{19}), would have been
inconsistent.

Thus, in conclusion, what has been shown here is that by imagining the
existence of many fermions at the edge (physically justified because of
the reasons mentioned at the beginning) and performing the non-Abelian
bosonisation of the corresponding edge
fermions, one can derive the Hall conductivity of the theory by looking at the
anomaly that is forced  on the system even before imposing chirality.   Having
done this, we can
then  impose chirality constraints, and it turns out that the equations
of motion are consistent with these constraints provided the gauging was done
suitably. In particular, this latter step also provided a physical
justification for the anomaly in the case when the charge $Q$ was central.
In this particular case, we explicitly check that our analogue of Jain's model
gives for the
Hall conductivity, just as in Jain's work, fractions with odd denominators.  It
is also clear from this
analysis that the generic feature of quantization of Hall conductivity can
be explained without requiring specific assumptions regarding the kind of
interactions present between the electrons because the infrared limit of a
certain class of theories gives rise to one whose conductivity is quantized.

\noindent
{\bf Acknowledgements}
\nopagebreak

I would like to thank A.P.Balachandran for suggesting this problem and
providing many useful suggestions throughout this work.  I would also like to
thank E.Ercolessi, D.Karabali, G.Landi, A.Momen, A.Qamar, B.Sathiapalan and
P.Teotonio for
many discussions.  I would also like to thank D.Karabali for removing a
confusion regarding the factor that appears in (\ref{18}). This work was
supported by a grant from DOE under contract number DE-FG02-85ER40231.

\end{document}